# A comparison of CME expansion speeds between solar cycles 23 and 24

**Fithanegest K Dagnew[1, 2, 3], Nat Gopalswamy[2] and Solomon B Tessema[1]**

[1]Ethiopian Space Science and Technology Institute (ESSTI), Entoto Observatory and Research Center (EORC), Addis Ababa, Ethiopia

[2] NASA Goddard Space Flight Center, Greenbelt, MD, USA

[3]The Catholic University of America, Washington DC, USA

E mail: fithanegest.k.dagnew@nasa.gov, dagnewfitha10@gmail.com

**Abstract**. We report on a comparison of the expansion speeds of limb coronal mass ejections (CMEs) between solar cycles 23 and 24. We selected a large number of limb CME events associated with soft X-ray flare size greater than or equal to M1.0 from both cycles. We used data and measurement tools available at the online CME catalog (https://cdaw.gsfc.nasa.gov) that consists of the properties of all CMEs detected by the Solar and Heliospheric Observatory's (SOHO) Large Angle and Spectrometric Coronagraph (LASCO). We found that the expansion speeds in cycle 24 are higher than those in cycle 23. We also found that the relation between radial and expansion speeds has different slopes in cycles 23 and 24. The cycle 24 slope is 45% higher than that in cycle 23. The expansion speed is also higher for a given radial speed. The difference increases with speed. For a 2000 km/s radial speed, the expansion speed in cycle 24 is ~48% higher. These results present additional evidence for the anomalous expansion of cycle 24-CMEs, which is due to the reduced total pressure in the heliosphere.

## 1. Introduction

The expansion speed ($V_{exp}$) of a CME is the rate at which the lateral dimension of a CME at its widest part changes with time. The radial speed ($V_{rad}$) is the speed of the CME nose which has an empirical relationship with $V_{exp}$ [1]. Schwenn et al. [2-3] define the expansion speed as the speed at which the CME expands in a direction perpendicular to its direction of propagation. The expansion speed serves as a proxy for the radial speed of all types of CMEs. The radial speed of CMEs originating close to the disk center of the Sun cannot be measured accurately because of projection effects.

Plunkett et al. [4] observed that for limb events, the cone angles (the angle between the outer edges of the opposing flanks of limb CMEs) of CME expansion and the shapes of expanding CMEs are usually well maintained. The CME shapes remained self-similar throughout the LASCO field of view. In other words, the ratio between the lateral expansion and radial propagation is constant. The shapes of most CMEs correspond with almost perfect circular cross section [5-6]. Although the sky plane speed varies considerably with the viewing angle, the expansion speed is independent of the direction of motion of the CME relative to the viewer's line of sight. The lateral CME expansion speed is the only parameter that can be uniquely measured for any CME, be it on the limb or pointed along the Sun-Earth line, on the front or back side [2-3].

It is well known that the state of the heliosphere in cycle 24 is considerably weak. There is a significant drop in the density, magnetic field, total pressure, and Alfven speed in the inner heliosphere as a result of low solar activity [7-8]. CME flux ropes expand as they propagate in the interplanetary space because the ambient pressure decreases with distance from the Sun [9]. The total pressure of the ambient medium at 1AU in cycle 24 was ~38% smaller than that in cycle 23 [10-11]. The reduced pressure allows CMEs to expand more and consequently become wider in cycle 24 for a given speed. Gopalswamy et al. [10] studied the speed and angular width of a large number of limb CMEs from solar cycles 23 and 24. The CME selection was based on soft X-ray flare size ($\geq$ C3.0) and source location (within 30° of the limb). They found similar correlation coefficients between the speed and angular width in both cycles 23 and 24 but a steeper regression line in cycle 24 compared to cycle 23. They reported that for a given CME speed, cycle 24 CMEs are significantly wider than cycle 23 CMEs indicating anomalous expansion which is caused by the weak state of the heliosphere. CME width typically increases over the first few solar radii, and then stabilizes to a constant value [8]. It has been suggested that CME flux ropes attain pressure balance at larger heliocentric distances in cycle 24 compared to cycle 23 [12].

Dal Lago et al. [1] studied 57 limb CMEs and obtained an empirical relationship Vrad = 0.88 Vexp, but did not consider the width dependence. Gopalswamy et al. [13] derived a theoretical relation $V_{rad}$ = f(w) $V_{exp}$, where w is the half width of the CME. They considered three cone models and found that for the full ice-cream cone model, f (w) = 1/2 (1+cotw), which provides the best fit to observations. Michalek et al. [14] studied 256 limb CMES from solar cycle 23 and obtained an empirical relationship between the radial and expansion speed as $V_{rad}$ = 1.17 $V_{exp}$. They reported that their result agrees with the full-cone model [13].

In this study, we report on a comparison of the expansion speeds of limb CMEs between solar cycles 23 and 24 and its implications for the state of the heliosphere.

## 2. Observations

We used data and measurement tools available at the on line (https://cdaw.gsfc.nasa.gov) CME catalog within the Solar and Heliospheric Observatory (SOHO) Large Angle and Spectrometric Coronagraph (LASCO) C2 and C3 field-of-view. We used JavaScript movies that combine LASCO images with GOES X-ray light curves. The LASCO C2 and LASCO C3 difference movies (c2rdif_gxray.html and c3rdif_gxray.html) are used to obtain the lateral width of the propagating CMEs and also to confirm the source locations. We selected a large number of limb CME events associated with soft X-ray flare size $\geq$ M1.0 from solar cycles 23 and 24 so that we can exclude the uncertainties in CME identification for weak flares. We considered 170 CMEs (85 from each cycle) which occur close to the solar limb (within $30^0$ from the limb: It is a subset of the events used in Gopalswamy et al. [10]). This is mainly to reduce projection effects in the measurement of speed and angular width. CMEs close to the solar limb are free from projection effects. The observations for cycle 23 cover the period from November 1997 to November 2003 and for cycle 24 from October 2011 to May 2017. The CME speeds ranged from ~150 km/s to >3000 km/s. There were 31 full halo CMEs (CME width, W= $360^0$) in cycle 24 and 24 in cycle 23. The number of partial halo CMEs (120° $\leq$W< 360°) were similar (28 in cycle 24 and 32 in cycle 23). The number of regular CMEs (W<120°) were also similar (26 in cycle 24 and 29 in cycle 23).

We took the radial speeds from the on line SOHO/LASCO (https://cdaw.gsfc.nasa.gov) CME catalog. The sky-plane speeds were obtained using a linear fit to the height time measurements made at the measurement position angle (MPA) where the CME seems to move the fastest [15]. Most of the radial speeds are the same as the sky plane speeds because of the minimal projection effects. For very few CMEs (with CMD around 60-65 degrees, we performed projection correction).

## 3. Methodology

We measured the expansion speed as the rate of change of the lateral width of the main body. The main body of the CME is thought to be the flux rope part of a CME, while the whole CME includes the shock part [16]. We obtained the angular and lateral widths of the CMEs observed by SOHO/LASCO. The angular widths of the CMEs may differ from the measurements in the SOHO/LASCO CME catalog (https://cdaw.gsfc.nasa.gov, [13-17]). The catalog measures the whole CME including the disturbance surrounding the flux rope, so the difference can be appreciable for very fast CMEs because they drive shocks. We considered the widest lateral dimension of the CME in successive frames of the propagating CME. We estimated the lateral extension by eye. We exclude the disturbances that appear as bright features around the flanks of the CME. While estimating the widest lateral extent, we have viewed movies of both direct and running difference images. Measurement continues until the angular width stabilizes, after which the lateral distance increases at a constant rate.

The lateral width is obtained using the measurement tool available online as part of the SOHO/LASCO CME catalog. The average expansion speed is computed from a linear fit to the lateral width versus time data points. The expansion speed of the 170 CMEs is obtained in this way.

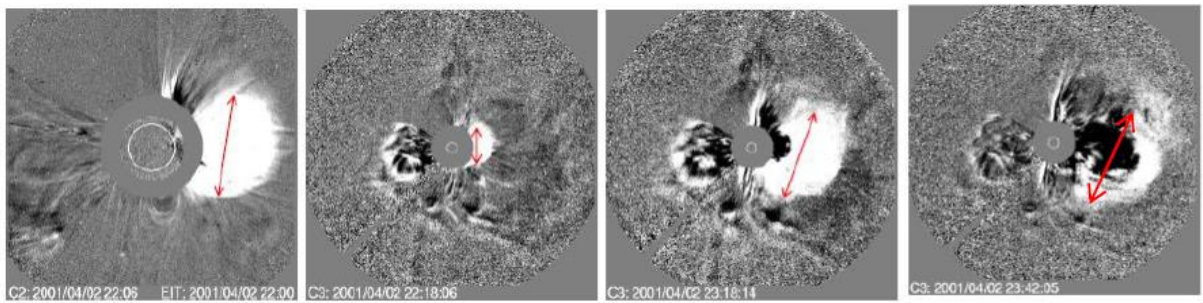

**Figure 1.** Illustration of the lateral width of the 2001 April 02 CME: The widest possible lateral end points of the consecutive frames are indicated by the double headed arrows.

*3.1. Lateral width measurement*

Figure 1 shows an example event for measuring the lateral width of a CME that occurred on 2001 April 02 at 22:20:07 UT. The angular width and the expansion speed obtained from our measurement are respectively $96^0$ and 2585 km/s. The sky-plane speed of the CME in the catalog is 2505 km/s. The ratio of the expansion speed to the radial speed obtained by our measurement for this CME is 1.03 ($V_{rad} = 0.97\ V_{exp}$). Using the measured angular width, we get f (w) = 0.95, so the expansion speed obtained from the full ice cream cone model [13] is 2637 km/s, which gives a ratio of 1.05 ($V_{rad} = 0.95\ V_{exp}$), in agreement with the measured expansion speed.

**4. Analysis and results**
We compare the distributions of the expansion speeds in cycles 23 and 24 using the scatter plots between the expansion speed and radial speed. We examine this relationship in cycles 23 and 24.

*4.1. Distribution of expansion speeds.*
Figure 2 compares the expansion speeds in cycles 23 and 24. The mean values of the expansion speeds in cycle 24 and cycle 23 are 1296 km/s and 1118 km/s respectively. The maximum values are 4010 km/s (cycle 24) and 3246 km/s (cycle 23). We see that 21% of the CMEs in cycle 24 have expansion speeds ≥ 2000 km/s while only 11% in cycle 23 have such speeds. The mean, median, standard deviation and maximum values of the expansion speeds in cycle 24 are higher than in cycle 23.

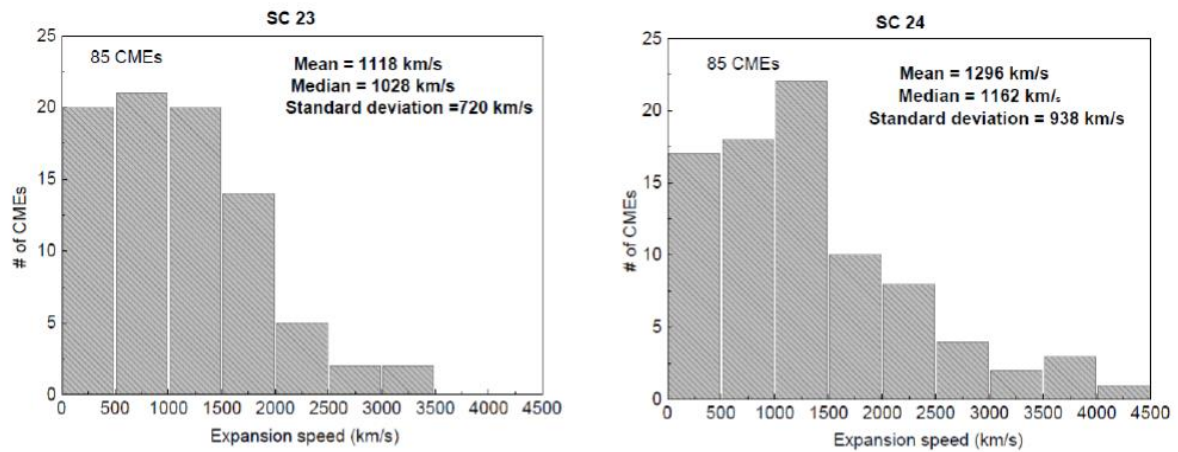

**Figure 2.** Comparison of expansion speeds in SC 23 and SC 24: The average expansion speed in cycle 24 is larger than in cycle 23.

*4.2. The Expansion speed - radial speed relationship*

Figure 3 shows the scatter plots between the expansion speed and radial speed of the 170 limb CMES in solar cycles 23 and 24. The regression lines and correlation coefficients (r) are shown on the plots. The slope and intercept values are indicated. The correlation coefficients are high (r > 0.9), which imply, a strong relationship between $V_{rad}$ and $V_{exp}$. The slope of the regression line in cycle 24 (1.48) is significantly higher than that in cycle 23 (1.03). The steeper regression line in cycle 24 implies that for a given CME radial speed, the expansion speed in cycle 24 is higher than that in cycle 23.

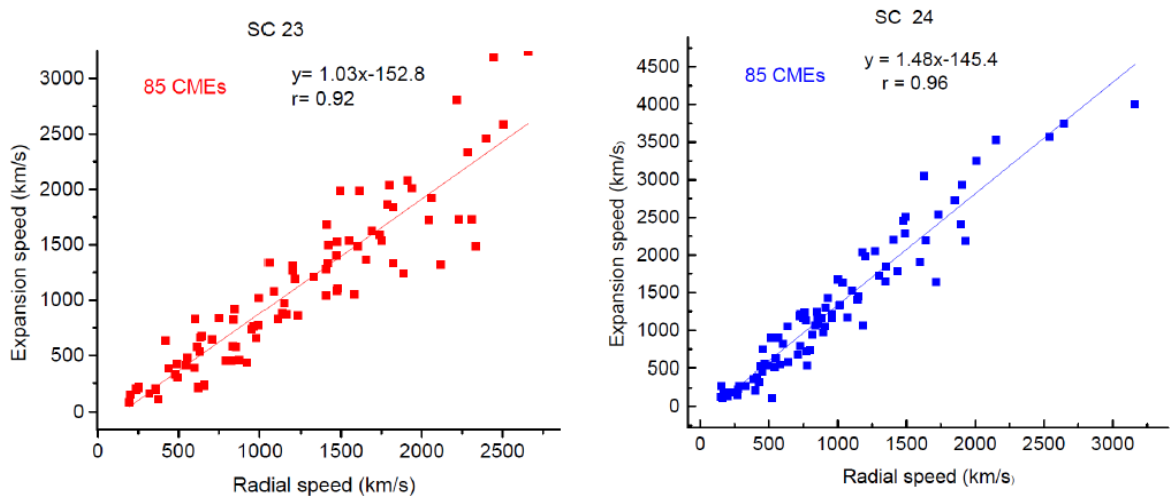

**Figure 3.** Scatter plots between the expansion speed and radial speeds of 170 limb CMEs in solar cycles 23(left) and 24 (right): The red and blue solid lines represent linear fit to the data points. The slope of the regression line and the correlation coefficients are shown on the plots.

The relationship between the radial and expansion speed obtained from our work is $V_{rad} = 0.97\ V_{exp}$ for cycle 23 and $V_{rad} = 0.68\ V_{exp}$ for cycle 24. Dal Lago et al. [1] obtained the relationship $V_{rad} = 0.88\ V_{exp}$ and Michalek et al. [14] obtained $V_{rad} = 1.17\ V_{exp}$. The theoretical relation derived by Gopalswamy et al. [13] is given by $Vrad = f(w)\ Vexp$ (w is the half width of the full ice-cream cone CME model, $f(w) = 1/2\ (1+cotw)$). According to this relationship: $f(w) = 0.97$ (cycle 23) and $f(w) = 0.68$ (cycle 24) corresponding to cone angles of $46.5^0$ and $70^0$, respectively. These imply an average width of $93^0$ (cycle 23) and $140^0$ (cycle 24) for the 85 CMEs in each cycle. The mean values of the width obtained from our measurement are $80^0$ and $114.5^0$ for cycles 23 and 24, respectively. The results we obtained are closer to this relationship.

Figure 4 shows a comparison of the slopes of the regression lines between cycles 23 and 24. The difference in the slopes increases with speed. The cycle 24 slope is 45% higher than that in cycle 23. The expansion speed is higher for a given radial speed. For a 2000 km/s radial speed, the expansion speed in cycle 24 is ~48% higher.

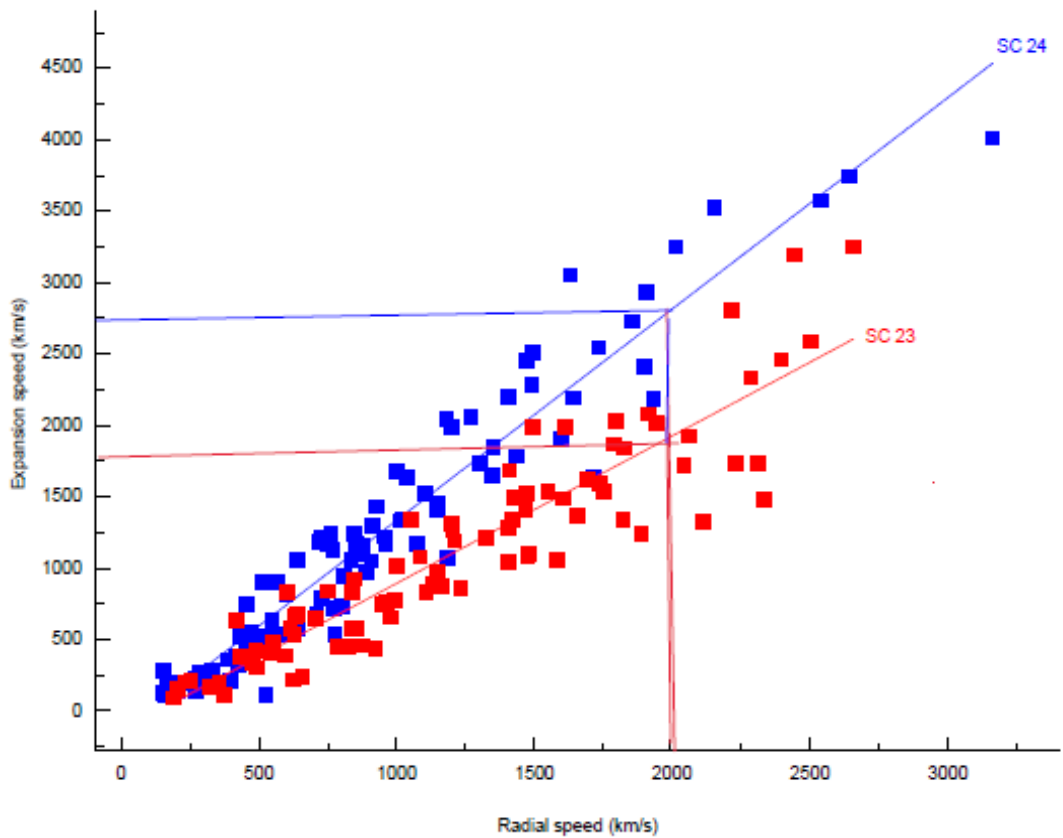

**Figure 4.** Comparison of the slopes of the regression lines in cycles 23 and 24. The vertical and horizontal lines mark the slope difference between the two cycles for a radial speed of 2000 km/s.

The scatterplots between CME speed (V) and angular width (W) by Gopalswamy et al. [10] show that cycle 24 regression line is steeper by 46% and for a speed of 1000 km/s and cycle 24 CMEs are wider by ~38%. This result is in agreement with our work: the average value of the measured width in cycle 24 is ~ 43% higher than in cycle 23. The angular width differences are proportional to the expansion speed differences. The expansion speed difference is another manifestation of the different interaction between the heliosphere and CMEs in the two cycles.

## 5. Summary and conclusion

We compared the expansion speeds of CMEs in solar cycles 23 and 24 for 170 limb CMEs (85 from each cycle). The histogram and statistical distributions show higher values of the expansion speed in cycle 24 as compared to cycle 23. The slopes of the regression lines in the scatter plot between the radial and expansion speeds are different for the two cycles. Cycle 24 slope is 45% higher than in cycle 23. The expansion speed is higher for a given radial speed. The difference increases with speed. For a 2000 km/s radial speed, the expansion speed in cycle 24 is 48% higher. The average value of the measured width in cycle 24 is significantly higher than in cycle 23. The expansion speed-radial speed relationship obtained from our work is $V_{rad} = 0.97\ V_{exp}$ and $V_{rad} = 0.68\ V_{exp}$ for cycles 23 and 24, respectively. This result is close to the theoretical relationship of the full cone model [13].

Given the previous finding from Gopalswamy et al. [10] and considering the higher slopes in cycle 24 and the tendency for significant increase in expansion speed values obtained from this work, we can say that the statistically significant difference can be attained with larger sample events. The overall results of this work present additional evidence for the anomalous expansion of cycle 24-CMEs, which is due to the reduced total pressure in the heliosphere.

**Acknowledgments**

We acknowledge NASA's open data policy in using SOHO/LASCO and STEREO data. This work is supported by NASA's Living with a Star program. FD thanks the Ethiopian Space Science and Technology Institute and the Mekelle University for partial financial support.

.